\documentclass[conference]{IEEEtran}
\IEEEoverridecommandlockouts
\usepackage{cite}
\usepackage{amsmath,amssymb,amsfonts}
\usepackage{graphicx}
\usepackage{textcomp}
\usepackage{xcolor}
\usepackage{algorithm,algpseudocodex}
\def\BibTeX{{\rm B\kern-.05em{\sc i\kern-.025em b}\kern-.08em
    T\kern-.1667em\lower.7ex\hbox{E}\kern-.125emX}}
\usepackage{geometry}
\begin{document}

\title{On Efficient Polyphase Network Implementation
Using Successive Vector Approximation
\thanks{This study was financed in part by the Coordena\c{c}\~{a}o de Aperfei\c{c}oamento de Pessoal de N\'{i}vel Superior - Brazil (CAPES) - Finance Code 001, and by CNPq and FAPERJ
both Brazilian research councils. Luiz F. da S. Coelho is a CNPq scholarship holder - Brazil. This work was partially supported by the ANR under the France 2030 program, grant NF-PERSEUS ANR-22-PEFT-0004.}
}

\author{\IEEEauthorblockN{Luiz F. da S. Coelho\IEEEauthorrefmark{1}\IEEEauthorrefmark{2}, Didier Le Ruyet\IEEEauthorrefmark{1}, and Paulo S. R. Diniz\IEEEauthorrefmark{2}}
\IEEEauthorblockA{\IEEEauthorrefmark{1}CEDRIC-CNAM, Paris, France\\
\IEEEauthorblockA{\IEEEauthorrefmark{2}Electrical Engineering Program, COPPE/UFRJ, Rio de Janeiro, Brazil}
Email: \{luizfelipe.coelho, diniz\}@smt.ufrj.br, didier.le\_ruyet@cnam.fr}}

\maketitle

\begin{abstract}
In this work, we explore an energy-efficient implementation of the polyphase network for a filter bank multicarrier (FBMC) system. The network is approximated using a greedy algorithm based on matching pursuits (MP) that converts the numerical representation directly from floating point to sum of signed powers of two (SOPOT), which is key for a multiplierless implementation. We compare this technique with other state-of-the-art methods for designing multiplierless hardware, and show that our technique achieves superior performance with similar computational complexity.
\end{abstract}


\section{Introduction}

Next-generation wireless systems aim to cover a wide variety of application scenarios, ranging from the connectivity of thousands of devices with low power consumption to higher data rates, high mobility communications, and low-complexity integrated sensing and communication (ISAC) operating in the GHz range \cite{NuriaISAC2024}. Within this context, waveforms play a crucial role, and their proper design is essential for coping with various challenges and providing more robust and efficient communications. Addressing the different criteria, including mitigating intersymbol interference, out-of-band radiation, computational cost, and robustness to a doubly selective channel, is a very challenging engineering problem. 

Orthogonal Frequency Division Multiplexing (OFDM), the standard waveform used in most modern communications systems, does not satisfy all those requirements, and different filtered waveforms have been proposed, including Filtered OFDM \cite{abdoli2015filtered} and  Orthogonal Time-Frequency Space Modulation based on Zak-transform (Zak-OTFS) \cite{mohammed2025zakotfscpofdm}.

Offset Quadrature Amplitude Modulation Filter Bank Multi-Carrier (OQAM-FBMC) system is an alternative to mitigate the high OOB emissions inherent to OFDM \cite{siohan2002analysis}. In OQAM-FBMC, since the subcarrier filtering process generates interference on the imaginary coefficients, the imaginary and the real parts are transmitted separately with a shift of half of the symbol period using OQAM. Power management is crucial in many communication systems, such as IoT, embedded devices, or power-limited devices, and the polyphase network (PPN) computation is a cost in OQAM-FBMC that can be reduced~\cite{nadal2018design}.


Multiplierless hardware design is a known technique for reducing the computational burden of multiplications in power-limited applications. These designs can be implemented in the form of look-up tables, where the inputs are directly mapped to the output without requiring numerical computation, or based on quantization, where bit-shift operators are applied instead. The former alleviates the computational burden by utilizing memory to store the mappings, and the latter modifies the numerical representation to perform fewer shifts. Both methods achieve approximated results, as not all numbers can be stored, and quantization is performed to reduce the number of shifts.

The vector quantization method approximates the vector elements into the so-called sums of signed powers of two (SOPOT), a numerical representation that employs signed powers of two (SPT) to reduce the number of necessary shifts in the multiplication.
While the conventional fixed-point representation uses the bit plane, or the set of possible powers of two, to allocate only zeros and ones, the SOPOT enables the extra possibility of a minus one. This creates a ternary numerical representation that reduces the number of active bits, allowing for more efficient exploration of the bit plane.

SOPOT has been successfully used in the design of finite impulse response (FIR) filters~\cite{dasilva2014fir}, and adaptive filters~\cite{coelho2022adaptive}. Moreover, in communication systems, SOPOT representation has been used for reduced complexity short prototype filters in OQAM-FBMC implementations~\cite{nadal2018design}, where the approximation is performed element by element. The main contributions of this work are:
\begin{itemize}
    \item The Signed Digit Loading (SDL), a simple greedy algorithm for approximating vectors into a SOPOT representation.
    \item A performance comparison, in terms of bit error rate (BER) and OOB radiation, between systems using the state-of-the-art multiplierless implementation, based on canonical signed digit (CSD), and the proposed vector approximation methods.
\end{itemize}
It is shown that the prototype filter of a PPN approximated using the proposed methods outperforms the state-of-the-art element-by-element approximation in all tested scenarios.

The remainder of the paper is organized as follows. Section~II presents the system model. Section~III describes the different proposed SOPOT approximation algorithms. Section~IV provides simulation results considering the PHYDYAS prototype filter. Finally, Section~V concludes the article.

\section{System Model}

  In OQAM-FBMC systems, the transmitted symbols $a_{k,n}$ are 
    Pulse Amplitude Modulation (PAM) modulated. According to the OQAM transmission, $d_{k,n}=a_{k,n}e^{j\phi_{k,n}}$ where $a_{k,n}$ are real symbols whose mean energy $\sigma^2_a$ is half of that of the QAM symbols' energy $\sigma^2_d$. $\phi_{k,n}$ is a phase term commonly defined in the literature as \cite{siohan2002analysis,zakaria2012novel}
    \begin{equation}
	    \phi_{k,n}=\frac{\pi}{2}(k+n)-\pi kn.
    \end{equation}
    Besides, the in-phase and quadrature components of the QAM symbols are staggered by half a symbol period, \textit{i.e.}, $T/2$.   

    Let $\bf g$ be a vector representing a prototype filter, such that
    \begin{align}
        {\bf g} = \begin{bmatrix}
            g[0] & g[1] & \cdots & g[OK-1]
        \end{bmatrix}^T,
    \end{align}
    where $g[m]$ is the $m$-th element of the vector, ${\bf g} \in \mathbb{R}^{OK}$, $O$ is the overlapping factor, indicating the number of overlapping information blocks, $K$ is the number of subcarriers, and ${\bf g}^T{\bf g} = 1$. The discrete-time transmitted signal in OQAM-FBMC can be written as \cite{pinchon2004design,zakaria2012novel}
    \begin{equation}\label{eq: tx_OQAM-FBMC1}
	    s[m] = \sum_{k=0}^{K-1}\sum_{n=0}^{N-1}d_{k,n}g\left[m-nK/2\right]e^{j\frac{2\pi}{K}km},
    \end{equation}

    The phase offset inserted by $\phi_{k,n}$ ensures that the frequency-time neighboring PAM symbols are in quadrature with each other.
    
    Defining $g_{k,n}[m]$ as
    \begin{equation}
        g_{k,n}[m] = g\left[m-nK/2\right]e^{j\frac{2\pi}{K}km}e^{j\phi_{k,n}},
    \end{equation}
    we can rewrite equation (\ref{eq: tx_OQAM-FBMC1}) as follows

 \begin{equation}\label{eq: tx_OQAM-FBMC}
	    s[m] = \sum_{k=0}^{K-1}\sum_{n=0}^{N-1}a_{k,n}g_{k,n}\left[m \right],
    \end{equation}

Assuming a noiseless and distortion-free channel, the demodulated symbol over the $k$th
subcarrier and the $n$th instant can be determined using the inner product of $s[m]$ and
$g_{k,n}[m]$:

\begin{align}
r_{k,n} &= \langle s, g_{k,n} \rangle 
= \sum_{m=-\infty}^{+\infty} s[m]\, g_{k,n}^*[m] \\
&= \sum_{k'=0}^{K-1} \sum_{n'=0}^{N-1} a_{k',n'} 
\sum_{m=-\infty}^{+\infty} g_{k',n'}[m]\, g_{k,n}^*[m] \notag \\
&=a_{k,n}+ \sum_{(k',n') \neq (k,n)} 
\sum_{m=-\infty}^{+\infty} g_{k',n'}[m]\, g_{k,n}^*[m]
\end{align}
The transmitted PAM symbol $a_{k,n}$ can be recovered by retrieving the real part of the demodulated signal $r_{k,n}$:
\begin{equation}
    \tilde{a}_{k,n}= \Re \{r_{k,n} \} = a_{k,n}+ I_{k,n}
\end{equation}
where $I_{k,n}$ is the residual interference. the variance of $I_{k,n}$ can be obtained as follows:
\begin{equation}
 \sigma^2_I= \sum_{(k,n)\neq (0,0)} \Bigg| \Re \Bigg \{ \sum_{m=-\infty}^{+\infty} g_{0,0}[m] g_{k,n}^*[m] \Bigg \} \Bigg|^2
\end{equation}

\section{SOPOT Approximation}

When dealing with numerical computation, multiplication is a known cost that should be considered in a system's power budget. Floating-point multiplication is costly and, in most cases, unnecessarily precise. Achieving approximate results via quantization is a known technique for reducing the computational burden and achieving \textit{good enough} results in most applications. In SOPOT, a number is represented as the sum of a limited number of signed powers of two (SPT); this alleviates the computational burden of the multiplication process, as each SPT operates as a bit shift.

Let $v \in \mathbb{R}$ be a floating-point value to be represented in SOPOT, and $\mathcal{S}=\{-1,0,1\}$ a set containing three elements. We employ symbols $c_i \in \mathcal{S}$ to indicate the presence of an SPT in the depth $i \in \mathbb{Z}$ of the bit plane, where higher values of $i$ indicate a deeper bit plane. The value is represented as
\begin{equation}
    v = \sum_i c_i 2^{-i},
\end{equation}
where the pair $(c_i, 2^{-i})$ is denoted an SPT. Without restrictions on the sum, the numerical representation is limitless in terms of precision. However, the infinite sum is impractical and undesirable for operations with reduced complexity. Hence, an approximation is achieved by limiting the number of SPTs.

The SOPOT numerical representation can be mathematically modeled as the following optimization problem
\begin{equation}
    \min_{\bf c} \quad \left| v - {\bf p}^T {\bf c} \right| \quad \textrm{s.t.} \quad || {\bf c} ||_0 \leq M_{\textrm{max}}, \label{eq:opt}
\end{equation}
where ${\bf p} \in \mathbb{R}^{B_{\textrm{max}}+1}$ is a vector containing $B_{\textrm{max}}$ powers of two, ${\bf c} \in \mathcal{S}^{B_{\textrm{max}}+1}$ is a sparse vector containing the symbols $c_i$, $M_{\textrm{max}}$ is the maximum number of SPTs, $|\cdot|$ is the absolute value operator, and $||\cdot||_0$ is the $\ell_0$-norm of a vector, which returns its number of nonzero elements. The vector $\bf p$ is the basis for the SOPOT, and is usually defined as
\begin{equation}
    {\bf p} = \begin{bmatrix}
        2^0 & 2^{-1} & \cdots & 2^{-B_{\textrm{max}}}
    \end{bmatrix}^T,
\end{equation}
for $|v| \leq 1$. Therefore, $B_{\textrm{max}}$ is the deepest bit plane an SPT can be allocated, and the product ${\bf p}^T{\bf c}$ results in the approximated value $\hat{v}$.

One notices that Equation \eqref{eq:opt} can have multiple solutions with different values for $||{\bf c}||_0$, which hinders the search for an optimal solution. A crucial solution is the CSD representation algorithm~\cite{diniz2010digital}, which begins with a 2's complement numerical representation (fixed-point) and reallocates SPTs across the bit planes. The CSD representation reduces the average number of active powers of two from $50\%$ in a 2's complement representation, to $30\%$ of the total wordlength~\cite{coelho2022adaptive}, which is an essential reduction for multiplierless implementations.

The CSD algorithm is limited by two critical factors: (i) it operates over scalars, and (ii) the active power of two reductions is limited by the wordlength of the $2$'s complement quantization. These limitations are significant when dealing with vectors, as certain elements might benefit from having different quantities of SPTs.
Inspired by the bit loading method, we propose a greedy algorithm called Signed Digit Loading, which is capable of allocating SPTs over a vector, reducing the reconstruction error. Moreover, in~\cite{dasilva2014fir}, an algorithm based on matching pursuits for vector quantization is proposed to approximate vectors into a SOPOT representation, called Matching Pursuits with Generalized Bit Planes (MPGBP).

\subsection{Canonical Sign Digit}

The CSD algorithm, presented in Algorithm~\ref{alg:csd}, starts from a 2's complement fixed-point quantized number and represents it using a sparsification algorithm that maximizes the number of zero elements in $\mathcal{I}$. Let ${\bf c}_{\textrm{2's}}$ be a $B$-length vector representing the number $\tilde{v}$ in 2's complement, such as
\begin{align}
    \tilde{v} = -c_{\textrm{2's}}(0) + c_{\textrm{2's}}(1) 2^{-1} + \cdots + c_{\textrm{2's}}(B-1) 2^{-(B-1)},
\end{align}
where $c_{\textrm{2's}}(b) \in \{0, 1\}$ is the $b$-th element of ${\bf c}_{\textrm{2's}}$.

\begin{algorithm}[hbt!]
    \caption{CSD for SOPOT representation} \label{alg:csd}
    \begin{algorithmic}
        \Require ${\bf c}_{\textrm{2's}}$
        \State $c_{\textrm{2's}}(-1) = c_{\textrm{2's}}(0)$, $c_{\textrm{2's}}(B+1) = \delta(B+1) = 0$
        \For{$i = B, B-1, \dots, 0$}
        \State $\theta(i) = c_{\textrm{2's}}(i) \oplus c_{\textrm{2's}}(i+1)$
        \State $\delta(i) = \overline{\delta(i)} \times \theta(i)$
        \State $c_{\textrm{CSD}}(i) = (1 - 2 c_{\textrm{2's}}(i-1))\delta(i)$
        \EndFor
        \State\Return ${\bf c}_{\textrm{CSD}}$
    \end{algorithmic}
\end{algorithm}

Algorithm~\ref{alg:csd} returns the vector ${\bf c}_{\textrm{CSD}}$, and we have,
\begin{align}
    \tilde{v} = {\bf p}^T {\bf c}_{\textrm{CSD}},
\end{align}
as the reconstruction of $\tilde{v}$.

\subsection{Signed Digit Loading}

Let ${\bf v} \in \mathbb{R}^N$ be a column vector, with $||{\bf v}||_\infty \leq 1$, a SOPOT approximation of this vector can be represented as follows
\begin{align}
    \hat{\bf v} = {\bf C}{\bf p},
\end{align}
where ${\bf C} \in \mathcal{S}^{N \times (B_{\textrm{max}}+1)}$ is a matrix containing the symbols that represent the SPT allocation over the bit planes of each vector element.

The matrix ${\bf C}$ can be filled using a greedy algorithm that, at each iteration, allocates an SPT according to the element of ${\bf v}$ with the highest absolute value. The SPT is then subtracted from ${\bf v}$, resulting in a residue that is used in the following iteration. Algorithm~\ref{alg:sdl} describes this process.

\begin{algorithm}[htb!]
    \begin{small}
        \caption{Signed Digit Loading}\label{alg:sdl}
        \begin{algorithmic}
            \Require ${\bf v} \in \mathbb{R}^N$, $M_{\textrm{max}}$
            \State $i\gets 0$, ${\bf r}_0 \gets {\bf v}$, $M \gets 0$, $\hat{\bf v} \gets {\bf 0}$, $B\gets 0$
            \While{$M < M_{\textrm{max}}$ \& $B\leq B_{\textrm{max}}$}
            \State $p_i \gets \textrm{find}\left(|{\bf r}_i|, ||{\bf r}_i||_\infty\right)$
            \If{$r (p_i) > 0$}
            \State $c_i \gets 1$
            \Else
            \State $c_i \gets -1$
            \EndIf
            \State $k_i \gets \left\lceil -\log_2 \left( \frac{4}{3}  \left| r(p_i) \right| \right) \right\rceil$
            \State $B \gets k_i$
            \If{$B \leq B_{\textrm{max}}$}
            \State $\hat{v}(p_i) \gets \hat{v}(p_i) + 2^{-k_i} c_i$
            \State $r(p_i) \gets r(p_i) - 2^{-k_i} c_i$
            \State ${\bf r}_{i+1} \gets {\bf r}_i$
            \State $M\gets M+1$
            \EndIf
            \State $i \gets i+1$
            \EndWhile
            \State \Return $\hat{\bf v}$, $(k_i , c_i, p_i) ~\forall i$
        \end{algorithmic}
    \end{small}
\end{algorithm}

In Algorithm~\ref{alg:sdl}, the operator $\textrm{find}({\bf a}, b)$ returns the position of the first $b$ in the vector ${\bf a}$, and the absolute value operator returns the absolute value of each element when the input is a vector. At each iteration, the algorithm returns a trio $(k_i, c_i, p_i)$, where $k_i \in \mathbb{Z}$ indicates the depth of the $i$-th SPT, corresponding to the sign $c_i \in \mathcal{S}$, and the $p_i$-th element of $\hat{\bf v}$. Hence, the matrix ${\bf C}$ is filled as the following
\begin{align}
    \{ {\bf C} \}_{m,n} = \begin{cases}
        c_i, &\quad m=p_i, n=k_i \\
        0, &\quad \textrm{otherwise}
    \end{cases}.
\end{align}

\begin{figure}[hbt!]
    \centering
    \includegraphics[width=\linewidth]{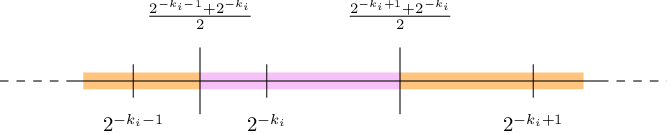}
    \caption{Illustration of the bounding region (in pink) for the bit plane depth allocation.}
    \label{fig:bounds}
\end{figure}

Moreover, we calculate $k_i$ as the nearest power of two neighbor, where we define the thresholding points as the middle point between $2^{-k_i-1}$ and $2^{-k_i}$ for the lower bound, and the middle point between $2^{-k_i+1}$ and $2^{-k_i}$ for the upper bound, as the following
\begin{align}
    \frac{2^{-k_i-1} + 2^{-k_i}}{2} \leq & \left|r(p_i)\right| < \frac{2^{-k_i} + 2^{-k_i+1}}{2} \\
    \frac{3}{4}2^{-k_i} \leq & \left|r(p_i)\right| < \frac{3}{4} 2^{-k_i+1} \\
    k_i \geq & -\log_2\left( \frac{4}{3} \left|r(p_i)\right| \right) > k_i - 1.
\end{align}
Therefore, one can calculate $k_i$ using the ceiling operator, as
\begin{align}
    k_i = \left\lceil -\log_2\left( \frac{4}{3} \left|r(p_i)\right|\right)\right\rceil.
\end{align}
These bounds are illustrated in Fig.~\ref{fig:bounds}, where the pink area delimits the region in which $|r(p_i)|$ receives $k_i$ as bit plane depth.

On one hand, this simple greedy algorithm iterates over the vector, and reaches an approximation using exactly $M_{\textrm{max}}$ SPTs for the numerical representation. On the other hand, a more general solution can be obtained using the MPGBP algorithm~\cite{dasilva2014fir}, where multiple elements of the residue can be utilized simultaneously.

\subsection{Matching Pursuits with Generalized Bit Planes}

Let ${\bf v} \in \mathbb{R}^N$ be a real-valued vector of length $N$, with $|| {\bf v} ||_{\infty} \leq 1$, i.e., maximum absolute value less or equal to $1$. An SOPOT approximated vector $\hat{\bf v}$ is represented as follows
\begin{equation}
    {\bf v} \approx \hat{\bf v} = \sum_{i=0}^{L-1} {\bf c}_{j_i} 2^{-k_i},
\end{equation}
where ${\bf c}_{j_i} \in \mathcal{S}^N$ is the $i$-th selected codeword from the dictionary $\mathcal{I}$, $k_i \in \mathbb{N}$ is the depth of the selected codeword, and $i \in \mathbb{N}$ is an index for the sum limited by $L \in \mathbb{N}$. The dictionary $\mathcal{I}$ is defined as all possible combinations of $N$-length vectors containing symbols from the set $\mathcal{S}^N$.

Let ${\bf r}_i \in \mathbb{R}^N$ be the residue from the $i$-th step of the algorithm, where the residue is defined as
\begin{equation}
    {\bf r}_i = {\bf r}_{i-1} - {\bf c}_{j_i} 2^{-k_i},
\end{equation}
and ${\bf r}_0 = {\bf v}$. The MPGBP algorithm greedily selects the codewords, ${\bf c}_{j_i}$, and the powers, $k_i$, that minimize the residue over the iterations. In the following, we describe these selections.

\subsubsection{Bit Depth Selection} The following equation defines the power, or the bit depth, for the $i$-th selected codeword
\begin{equation}
    k_i = \left\lceil - \log_2 \left( \frac{4 {\bf r}_{i-1}^T {\bf c}_{j_i} }{3 P} \right) \right\rceil,
\end{equation}
where $P \in \mathbb{N}$ defines the number of nonzero elements in the codeword. It is based on the nearest neighbor optimal quantization rule, and is the value that approximates the product ${\bf r}_{i-1}^T{\bf c}_{j_i}$ to $2^{-k_m}$~\cite{dasilva2014fir}.

\subsubsection{Codeword Selection} Instead of testing every possible codeword, \cite{dasilva2014fir} proposes a heuristic consisting of selecting $P$ elements with the greatest absolute values from the residue. These elements' signs and indices (position) fill the codeword ${\bf c}_{j_i}$, resulting in a vector with $P$ nonzero elements and $N-P$ zeros. It is also shown that $P = \lfloor \sqrt{N} \rfloor$ provides an approximation using the minimum number of SPTs. The MPGBP algorithm is presented in pseudocode for Algorithm~\ref{alg:mpgbp}.

\begin{algorithm}[htb!]
    \begin{small}
        \caption{MPGBP for vector approximation}\label{alg:mpgbp}
        \begin{algorithmic}
            \Require ${\bf v} \in \mathbb{R}^N$, $M_{\textrm{max}}$, and $B_{\textrm{max}}$
            \State $i \gets 0$, ${\bf r}_0 \gets {\bf v}$, $M \gets 0$, $B \gets 0$, $P \gets \left\lfloor \sqrt{N} \right\rfloor$, $\hat{\bf v} \gets {\bf 0}$
            \While{$M < M_{\textrm{max}}$ \& $B < B_{\textrm{max}}$}
            \State {\bf codeword search:} Sort indexes of ${\bf r}_i$ in decreasing order of the absolute value and store the first $P$ indexes in $\mathcal{I}_P$
            \State {\bf codeword update:}
            \State ${\bf c}_{j_i} \gets {\bf 0}$
            \For{$p \in \mathcal{I}_P$}
            \If{$r_i (p) > 0$}
            \State $c_{j_i}(p) \gets 1$
            \Else
            \State $c_{j_i}(p) \gets -1$
            \EndIf
            \EndFor
            \State {\bf residue update:}
            \State $k_i \gets \left\lceil -\log_2 \left( \frac{4}{3P} \sum_{p \in \mathcal{I}_P} \left| r_i(p) \right| \right) \right\rceil$
            \State $\hat{\bf v} \gets \hat{\bf v} + 2^{-k_i} {\bf c}_{j_i}$
            \State ${\bf r}_{i+1} \gets {\bf r}_i - \hat{\bf v}$
            \State $B \gets \max(k_i, B)$
            \State $M\gets \sum_i \sum_n |c_{j_i}(n)|$
            \State $i \gets i+1$
            \EndWhile
            \State \Return $\hat{\bf v}$, $(k_i , {\bf c}_{j_i}) ~\forall i$
        \end{algorithmic}
    \end{small}
\end{algorithm}

\subsubsection{Stopping Criteria} There are many possible stopping criteria for the MPGBP algorithm. Although we use $M_\textrm{max} \in \mathbb{N}$ as the maximum number of SPTs used in the vector's representation, and $B_\textrm{max} \in \mathbb{N}$ as the maximum bit plane depth, one can base the stopping criterion on the residue ${\bf r}_i$ evaluating the approximation. As the number of iterations grows, $\hat{\bf v}$ converges to ${\bf v}$, and the residue converges to ${\bf 0}$; this implies that $k_i$ grows with each iteration as smaller powers of two are necessary to reduce the residue further. Moreover, each iteration adds $P$ new SPTs to $\hat{\bf v}$; if the criterion $B_\textrm{max}$ is not attained, the number of iterations is $\lceil \frac{M_\textrm{max}}{P} \rceil$ resulting in $M = \lceil \frac{M_\textrm{max}}{P} \rceil P$ SPTs in the final approximation.

An essential difference between the CSD and the vector approximations is that the latter allocates resources (SPTs) to minimize the mean-squared residual difference between the approximated and the original vectors, and the former is an element-by-element bit-to-ternary transformation aiming at reducing the number of SPTs in the fixed-point numerical representation, i.e., it is a post-quantization process. Although both methods achieve the SOPOT representation, we show that the vector approximation is better suited for approximating prototype filters.

\section{Simulation Results}

\begin{figure}[hbt!]
    \centering
    \includegraphics[width=\linewidth]{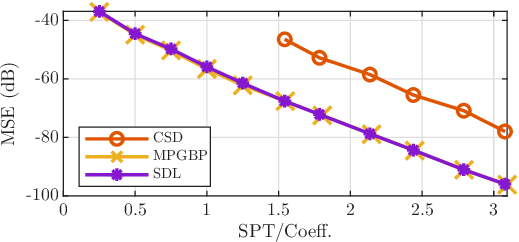}
    \caption{MSE for the PHYDYAS filter approximations, using the CSD, SDL, and MPGBP algorithms in different complexity levels.}
    \label{fig:mse}
\end{figure}

To evaluate how well each method approximates the prototype filter, $\bf g$, we calculate the mean squared error (MSE) between the original filter and the approximations. These results are presented in Fig.~\ref{fig:mse}. One notices that the MPGBP and the SDL approximations have, on average, a $20~\textrm{dB}$ gain in terms of MSE for approximations with similar complexity, i.e., approximations with an equivalent number of SPT per coefficient (or vector element). The CSD algorithm uses $3$, $4$, $5$, $6$, $7$, and $8$-bit fixed-point quantization, resulting in an average of $1.5$, $1.8$, $2.1$, $2.4$, $2.8$, and $3.1$ SPT/Coeff., respectively.

\begin{figure}[hbt!]
    \centering
    \includegraphics[width=\linewidth]{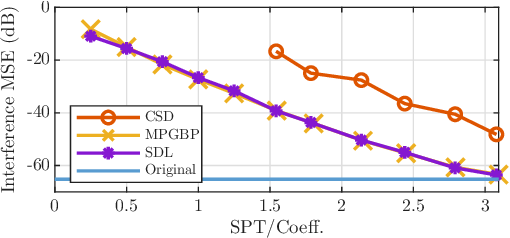}
    \caption{Interference MSE for the PHYDYAS filter approximations, using the CSD, SDL, and MPGBP algorithms in different complexity levels.}
    \label{fig:mse_interf}
\end{figure}

Moreover, to evaluate its performance in terms of interference cancellation, Fig.~\ref{fig:mse_interf} presents the calculated interference MSE for OQAM-FBMC systems using the approximated filters. One also observes a $20~\textrm{dB}$ gap in the interference MSE between the vector approaches (SDL and MPGBP) and CSD approximations. These results show that, despite being designed to minimize the MSE, the vector approximation algorithms also outperform the CSD in terms of residual interference.

We also compare the performance of the quantized system in terms of BER and OOB radiation. The overall system parameters are $M=128$ subcarriers over $N=64$ blocks of information; we use the PHYDYAS prototype filter, with an overlapping factor $K=4$ \cite{bellanger2010fbmc}. As the results of the SDL algorithm and the MPGBP present no significant difference, we continue the performance evaluation, showing results only for the SDL algorithm for the vector approximation method. We compare two systems with similar computational costs in a multiplierless configuration: the $4$-bit CSD and an approximation using the SDL with $1.8~\textrm{SPT/Coeff.}$. We also compare the system's performance for different levels of approximation using the SDL algorithm.

\begin{figure}[hbt!]
    \centering
    \includegraphics[width=\linewidth]{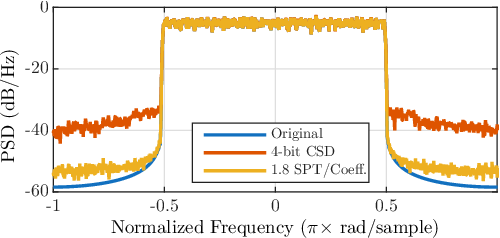}
    \caption{PSD of FBMC/OQAM systems with similar complexity; $4$-bit CSD obtained using the CSD algorithm, $1.8~\textrm{SPT/Coeff.}$ obtained using the SDL algorithm, and Original as an estimate for the system using a non-approximated prototype filter.}
    \label{fig:csd-vs-mpgbp}
\end{figure}

\begin{figure}[hbt!]
    \centering
    \includegraphics[width=\linewidth]{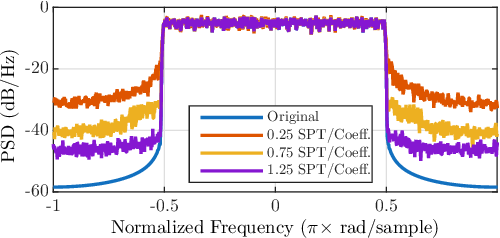}
    \caption{PSD of FBMC/OQAM systems with different approximation levels, using the SDL algorithm.}
    \label{fig:mpgbp-compare}
\end{figure}

To evaluate the OOB radiation, we generate an OQAM-FBMC signal illuminating only the $64$ central subcarriers with $4$-QAM symbols following an i.i.d. distribution and estimate the signal's power spectral density (PSD). These results are presented in Figs.~\ref{fig:csd-vs-mpgbp}~and~\ref{fig:mpgbp-compare}.

In Fig.~\ref{fig:csd-vs-mpgbp}, one observes that the $4$-bit CSD approximation results in an elevation gap of approximately $20~\textrm{dB}$ in the OOB PSD level when compared to the Original system. At the same time, its SDL counterpart presents results much closer to the non-approximated prototype filter. Moreover, Figure~\ref{fig:mpgbp-compare} illustrates the effect of varying approximation levels on the system's PSD; it is observed that coarser approximations, i.e., fewer SPT/Coeff., result in higher levels of OOB radiation.

\begin{figure}[hbt!]
    \centering
    \includegraphics[width=\linewidth]{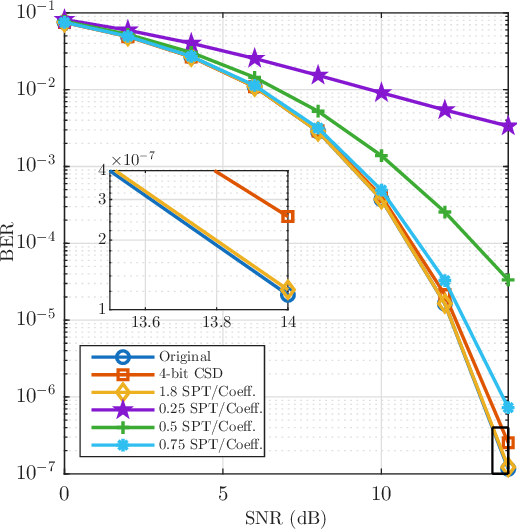}
    \caption{BER for FBMC/OQAM using a 4-QAM constellation, with approximations using the 4-bit CSD, and the SDL for different approximation levels.}
    \label{fig:ber-4qam}
\end{figure}

\begin{figure}[hbt!]
    \centering
    \includegraphics[width=\linewidth]{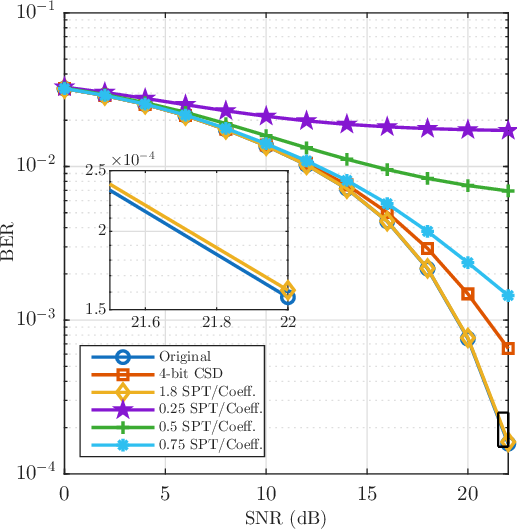}
    \caption{BER for FBMC/OQAM using a 64-QAM constellation, with approximations using the 4-bit CSD, and the SDL for different approximation levels.}
    \label{fig:ber-64qam}
\end{figure}

The BER is estimated in a Monte Carlo process using $10000$ samples, for systems using all $M=128$ subcarriers in a purely AWGN channel. Figs.~\ref{fig:ber-4qam}~and~\ref{fig:ber-64qam} present the BER results for systems using $4$-QAM and $64$-QAM constellations, respectively.

In Fig.~\ref{fig:ber-4qam}, we observe that both the 4-bit CSD and its SDL counterpart have a BER performance very similar to the original system, where the SDL slightly outperforms the CSD approximation. However, coarser SDL approximations have poor performance; in our experiments, the performance is maintained similar to the original, with MPGBP approximations down to an average of 1 SPT/Coeff. for systems using 4-QAM constellations.

Fig.~\ref{fig:ber-64qam} presents the results for systems using 64-QAM constellations. It shows that the 4-bit CSD is a considerably less effective approximation when dealing with more densely populated constellations. However, the SDL using $1.8$ SPT/Coeff. performs similarly to the original system.

These results demonstrate that vector approximation methods enable a better-performing multiplierless PPN than the state-of-the-art CSD algorithm. Although the CSD performs similarly in terms of BER in 4-QAM, the OOB radiation of the vector approach is significantly better, presenting a significant improvement even in less dense constellations.

\section{Conclusion}
This paper proposes a method to reduce the computational burden of filter bank multicarrier systems by employing two greedy algorithms: the SDL and the matching-pursuit-based MPGBP algorithm, to generate multiplierless solutions. The resulting FBMC transceivers nearly maintain their low OOB and BER while employing a small number of signed powers of two, leading to computationally efficient solutions paving the way to meet the challenges of demanding applications. The solutions enable low-energy and fast implementations when hardware constraints are imposed.

Future work focuses on designing fully SPT-based multicarrier transceivers and addressing their practical performance, considering overall complexity in relation to OOB and ISI. It is worth mentioning that the proposed work can be applied to other filtered multicarrier waveforms, including Fast Fourier Transform (FFT)-FBMC \cite{zakaria2012novel} or  2D-FFT-FB \cite{junior2023two}.

\bibliographystyle{IEEEtran}
\bibliography{ref}

\end{document}